\documentclass[aps,pra,superscriptaddress,twocolumn,nofootinbib]{revtex4-2}

\usepackage{amsfonts}
\usepackage{amsmath}
\usepackage[normalem]{ulem}
\usepackage{amssymb}
\usepackage{mathtools}
\usepackage{graphicx}
\usepackage{enumerate}
\usepackage{wrapfig}
\usepackage{color}
\usepackage{yhmath}
\usepackage{mathrsfs}
\usepackage{bbm}
\usepackage{commath}
\usepackage[hidelinks]{hyperref}
\hypersetup{
     colorlinks   = true,
     citecolor    = blue,
     linkcolor    = blue
}

\DeclarePairedDelimiterX{\relentrx}[2]{(}{)}{%
  #1\;\delimsize\|\;#2%
}
\newcommand{\relentr}{S\relentrx}

\usepackage{subfigure}
\usepackage{mathptmx}
\usepackage{times}
\usepackage{ulem}
\usepackage{lipsum}

\DeclareMathAlphabet{\pazocal}{OMS}{zplm}{m}{n}

\DeclareMathOperator{\tr}{tr}

\DeclareMathSymbol{\Alpha}{\mathalpha}{operators}{"41}
\DeclareMathSymbol{\Beta}{\mathalpha}{operators}{"42}
\DeclareMathSymbol{\Epsilon}{\mathalpha}{operators}{"45}
\DeclareMathSymbol{\Zeta}{\mathalpha}{operators}{"5A}
\DeclareMathSymbol{\Eta}{\mathalpha}{operators}{"48}
\DeclareMathSymbol{\Iota}{\mathalpha}{operators}{"49}
\DeclareMathSymbol{\Kappa}{\mathalpha}{operators}{"4B}
\DeclareMathSymbol{\Mu}{\mathalpha}{operators}{"4D}
\DeclareMathSymbol{\Nu}{\mathalpha}{operators}{"4E}
\DeclareMathSymbol{\Omicron}{\mathalpha}{operators}{"4F}
\DeclareMathSymbol{\Rho}{\mathalpha}{operators}{"50}
\DeclareMathSymbol{\Tau}{\mathalpha}{operators}{"54}
\DeclareMathSymbol{\Chi}{\mathalpha}{operators}{"58}
\DeclareMathSymbol{\omicron}{\mathord}{letters}{"6F}

\def\Xint#1{\mathchoice
{\XXint\displaystyle\textstyle{#1}}%
{\XXint\textstyle\scriptstyle{#1}}%
{\XXint\scriptstyle\scriptscriptstyle{#1}}%
{\XXint\scriptscriptstyle\scriptscriptstyle{#1}}%
\!\int}
\def\XXint#1#2#3{{\setbox0=\hbox{$#1{#2#3}{\int}$}
\vcenter{\hbox{$#2#3$}}\kern-.5\wd0}}

\def\dashint{\Xint-}

\makeatletter
\DeclareFontFamily{OMX}{MnSymbolE}{}
\DeclareSymbolFont{MnLargeSymbols}{OMX}{MnSymbolE}{m}{n}
\SetSymbolFont{MnLargeSymbols}{bold}{OMX}{MnSymbolE}{b}{n}
\DeclareFontShape{OMX}{MnSymbolE}{m}{n}{
    <-6>  MnSymbolE5
   <6-7>  MnSymbolE6
   <7-8>  MnSymbolE7
   <8-9>  MnSymbolE8
   <9-10> MnSymbolE9
  <10-12> MnSymbolE10
  <12->   MnSymbolE12
}{}
\DeclareFontShape{OMX}{MnSymbolE}{b}{n}{
    <-6>  MnSymbolE-Bold5
   <6-7>  MnSymbolE-Bold6
   <7-8>  MnSymbolE-Bold7
   <8-9>  MnSymbolE-Bold8
   <9-10> MnSymbolE-Bold9
  <10-12> MnSymbolE-Bold10
  <12->   MnSymbolE-Bold12
}{}

\newcommand{\ignore}[1]{}
\newcommand{\nobibentry}[1]{{\let\nocite\ignore\bibentry{#1}}}

\newcommand{\bibfnamefont}[1]{#1}
\newcommand{\bibnamefont}[1]{#1}
\newcommand{\ket}[1]{\left\vert#1\right\rangle}
\newcommand{\bra}[1]{\left\langle#1\right\vert}

\newcommand{\bea}{\begin{eqnarray}}
\newcommand{\eea}{\end{eqnarray}}

\newcommand{\Hint}{\pmb{V}}

\renewcommand*{\thefootnote}{\fnsymbol{footnote}}

\newcommand*{\blue}[1]{\textcolor{black}{#1}}

\begin{document}

\title{Accurate heat currents via reorganised master equation}

\author{Jonas Glatthard}
\affiliation{Department of Physics and Astronomy, University of Exeter, Exeter EX4 4QL, United Kingdom}

\author{Guillem Aznar-Menargues}
\affiliation{Instituto Universitario de Estudios Avanzados (IUdEA), Universidad de La Laguna, La Laguna 38203, Spain}
\affiliation{Departamento de Física, Universidad de La Laguna, La Laguna 38203, Spain}

\author{Jos\'{e} P. Palao}
\affiliation{Instituto Universitario de Estudios Avanzados (IUdEA), Universidad de La Laguna, La Laguna 38203, Spain}
\affiliation{Departamento de Física, Universidad de La Laguna, La Laguna 38203, Spain}

\author{Daniel Alonso}
\affiliation{Instituto Universitario de Estudios Avanzados (IUdEA), Universidad de La Laguna, La Laguna 38203, Spain}
\affiliation{Departamento de Física, Universidad de La Laguna, La Laguna 38203, Spain}

\author{Luis A. Correa}
\affiliation{Instituto Universitario de Estudios Avanzados (IUdEA), Universidad de La Laguna, La Laguna 38203, Spain}
\affiliation{Departamento de Física, Universidad de La Laguna, La Laguna 38203, Spain}
\affiliation{Department of Physics and Astronomy, University of Exeter, Exeter EX4 4QL, United Kingdom}
\email{lacorrea@ull.edu.es}

\begin{abstract}
The accurate characterisation of energy exchanges between nanoscale quantum systems and their environments is of paramount importance for quantum technologies, and central to quantum thermodynamics. Here, we show that, in order to accurately approximate steady-state heat currents via perturbative master equations, the coupling-induced reorganisation correction to the system's energy must be carefully taken into account. Not doing so, may yield sizeable errors, especially at low, or even moderate temperatures. In particular, we show how a `reorganised master equation' can produce very accurate estimates for the heat currents when the reorganisation energy is weak and one works with environments with a broad spectrum. Notably, such master equation outperforms its `non-reorganised' counterpart in the calculation of heat currents, at modelling dynamics, and at correctly capturing equilibration. This is so even if both types of equation are derived to the same order of perturbation theory. Most importantly, working with reorganised master equations does not involve additional complications when compared with alternative approaches. Also, invoking the secular approximation to secure thermodynamic consistency does not compromise their precision.
\end{abstract}

\maketitle

\section{Introduction}

The non-equilibrium steady state of a weakly dissipative open quantum system can be studied through master equations \cite{bp}. One may think, for concreteness, of a system coupled to various heat baths\footnote{In general, it could also be subjected to external driving \cite{alicki1979,geva1992a,alicki2018}.} at different temperatures $ T_i $. In particular, master equations allow for the characterisation of the stationary heat currents $ \blue{\dot{\mathcal{Q}}_i} $ flowing from each bath into the system. Provided that these abide by a relationship of the form \cite{alicki1979}
\begin{equation}\label{eq:clausius-inequality}
   \sum\nolimits_i \frac{\dot{\mathcal{Q}}_i}{T_i} \leq 0,
\end{equation}
which is analogous to the Clausius inequality, one may regard such non-equilibrium open system as a continuous thermodynamic cycle \cite{palao2001,kosloff2014}. By tuning system's parameters so as to change the direction and magnitude of the steady-state heat currents, one can switch between cooling and heat-engine operation modes, or seek to optimise energy performance according to some relevant figure of merit \cite{geva1992b,correa2014,whitney2014,feldmann2006,correa2015,abiuso2020}. One can even ask whether nanoscale thermodynamic devices are distinctively `quantum' in any measurable way, and whether their `quantumness' may be exploited in practice \cite{scully2011,raam2015,klatzow2019,gonzalez2019}. Theoretical progress on quantum thermodynamics has also motivated much experimental work, and this type of energy-conversion cycles have now been implemented on various platforms \cite{binder2018}, ranging from trapped ions \cite{rossnagel2016,lindenfels2019,maslennikov2019} to cold atomic gases \cite{brantut2013,zou2017}, quantum dots \cite{thierschmann2015}, or colour centres in diamond \cite{klatzow2019}. 

It thus seems clear that the characterisation of the stationary heat currents flowing through open systems is of paramount importance. However, this strongly depends on the master equation used to characterise them, which introduces some ambiguity. Indeed, working with master equations which are perturbative in the system--bath coupling, affords flexibility in their derivation \cite{winczewski2021,Potts_2021,lobejko202,correa2023}. Specifically, one has freedom to set the reference basis in which dissipative processes occur, by perturbing the microscopic Hamiltonian of the system \cite{correa2023}. Here, we show how choosing the basis of the system Hamiltonian renormalised by the various system--bath couplings, can yield substantially better master equations. Namely, we illustrate how the exact stationary heat currents \blue{in and out of linear bosonic baths}, the dissipative dynamics of the system, and its steady state may all be approximated much more closely than with alternative master equations, even when these are derived to the \textit{same order} in perturbation theory.  

This paper is structured as follows: In Sec.~\ref{sec:heat-currents} we address the issue of the definition of stationary heat currents within quantum thermodynamics. Next, in Sec.~\ref{sec:reorganised-master-equation} we introduce the `reorganised master equation' framework and discuss the influence of the choice of reference basis for dissipative (jump) processes on the definition of stationary currents. In Sec.~\ref{sec:examples} we illustrate our discussion with two simple examples: the quantum damped harmonic oscillator and the spin--boson model. Finally, in Sec.~\ref{sec:conclusions} we summarise and conclude.

\section{Steady-state heat currents}\label{sec:heat-currents}

To appreciate the underlying subtleties in the definition of the stationary heat currents across an open quantum system, we need to introduce some notation. Let the master equation approximating the reduced dynamics of the open system be
\begin{equation}\label{eq:generic-master-equation}
    \frac{d\pmb{\varrho}_S}{dt} = -i[\pmb{H}-\Delta\pmb{H},\pmb{\varrho}_S] + \mathcal{L}_{\pmb{H}}\,\pmb{\varrho}_S,
\end{equation}
where operators are denoted in boldface. The state of the system is here $\pmb{\varrho}_S$, $[\cdot,\cdot]$ stands for commutator, and the calligraphic symbol $ \mathcal{L}_{\pmb{H}} $ represents a dissipation super-operator (or \textit{dissipator}), responsible for excitation and decay processes in the basis of the reference system Hamiltonian $ \pmb{H} $. Since we deal with various independent heat baths at temperatures $ T_i $, we may always decompose $ \mathcal{L}_{\pmb{H}} = \sum\nolimits_i \mathcal{L}_{\pmb{H}}^{(i)} $. Finally, the Lamb-shift term $\Delta\pmb{H}$ is a dissipative correction to $ \pmb{H} $ (see Eq.~\eqref{eq:lamb-shift-term} below).

\blue{The underlying microscopic Hamiltonian for the open system has the form $\pmb{H}_\text{tot} = \pmb{H}_S + \pmb{H}_\text{int} + \sum\nolimits_i\pmb{H}_{B_i}$, where the system--bath coupling term has the form \begin{equation}\label{eq:H_int}
    \pmb{H}_\text{int} = \sum\nolimits_i \zeta_i\,\pmb{S}_i\otimes\pmb{B}_i,
\end{equation}
and $ \pmb{S}_i $ and $\pmb{B}_i$ are system and a bath operators, respectively. Here, the dimensionless parameters $\zeta_i$ captures the strength of each system--bath interaction\footnote{\blue{Here,  $\langle\pmb{B}_i\rangle=0$, but this can be enforced by absorbing the non-zero average into the system Hamiltonian \cite{winczewski2021}.\label{foot:first-order}}}. The bath is assumed to be initially uncorrelated from the system and in thermal equilibrium. Importantly, Eq.~\eqref{eq:generic-master-equation} assumes that the couplings are all small, and neglects any terms beyond second order in $\zeta_i$. As we shall see below, the reference Hamiltonian $\pmb{H}$ in Eq.~\eqref{eq:generic-master-equation} is not necessarily equal to $\pmb{H}_S$.}

In quantum thermodynamics, the dissipators $\mathcal{L}_{\pmb{H}}^{(i)}$ are most often assumed to have the Davies or Gorini--Kossakowski--Lindblad--Sudarshan (GKLS) form \cite{davies1974,lindblad1976,gks1976}; namely,
\begin{equation}\label{eq:lindblad-dissipator}
\mathcal{L}_{\pmb{H}}^{(i)}\,\pmb{\varrho}_S = \sum\nolimits_{\omega} \gamma_i(\omega)\,\left(\pmb{A}^{(i)}_{\omega}\,\pmb{\varrho}_S\,{\pmb{A}_\omega^{(i)}}^{\dagger} - \frac12\{ {\pmb{A}_\omega^{(i)}}^{\dagger}\,\pmb{A}_\omega^{(i)},\pmb{\varrho}_S\}\right).
\end{equation}
Here, the sum runs over all Bohr frequencies $\omega$ of the reference Hamiltonian $ \pmb{H} $, $\{\cdot,\cdot\}$ stands for anticommutator and $\gamma_i(\omega)$ is the rate of decay at frequency $\omega$ and temperature $ T_i $. These rates \blue{are calculated from the two-point correlation function of the bath coupling operators $\pmb{B}_i$ in the interaction picture (see Eq.~\eqref{eq:decay-rates} below) and} obey the detailed balance condition $\gamma_i(-\omega)/\gamma_i(\omega) = e^{-\omega/T_i}$ ($ \hbar = k_B = 1 $). The non-Hermitian operators $\pmb{A}_\omega^{(i)}$ characterise the bath-assisted transitions occurring at rate $\gamma_i(\omega)$, and are chosen so that $ \pmb{S}_i = \sum\nolimits_\omega \pmb{A}_{\omega}^{(i)} $ and $[\pmb{H},\pmb{A}_\omega^{(i)}] = - \omega\,\pmb{A}_\omega^{(i)}$. Also, in the case of GKLS dynamics, the Lamb-shift term satisfies $[\Delta\pmb{H},\pmb{H}] = 0$.

Although seemingly very specific, Eq.~\eqref{eq:lindblad-dissipator} emerges naturally under some basic thermodynamic requirements; namely, that the dynamics be fully \textit{contractive} towards a unique steady determined by temperature, and that no energy is stored at the system--baths interfaces \cite{dann2021}. Indeed, the dynamics generated by a GKLS equation has the key property \cite{alicki1979,spohn1978,bp}
\begin{equation}\label{eq:contractivity}
    \frac{d}{dt}\relentr{\pmb{\varrho}_S(t)}{\pmb{\tau}} \leq 0,
\end{equation}
where $ \relentr{\pmb{\rho}}{\pmb{\sigma}} \coloneqq \tr\,\pmb{\rho}(\log{\pmb{\rho}}-\log{\pmb{\sigma}}) $ is the quantum relative entropy between given states $ \pmb{\rho} $ and $ \pmb{\sigma} $, and $\pmb{\tau} = \lim_{t\rightarrow\infty}\pmb{\varrho}_S(t)$ stands for the steady state of the system. The relative entropy is non-negative and vanishes iff $\pmb{\rho} = \pmb{\sigma} $. Hence, it is frequently used as a notion of distance between states (even if it is non-symmetric). 

Eq.~\eqref{eq:contractivity} thus states that the system's `trajectory' takes it ever closer to its (non-equilibrium) steady state $\pmb{\tau}$. Since all the dissipators $\mathcal{L}_{\pmb{H}}^{(i)}$ have GKLS structure, the same can be said for each of them individually, i.e., if we were to `switch off' the couplings to all but the $i$-th bath in Eq.~\eqref{eq:generic-master-equation}, the resulting dynamics would still fulfill
\begin{equation*}
    \frac{d}{dt}\relentr{\pmb{\varrho}_S(t)}{\pmb{\tau}_i}\leq 0,
\end{equation*}
where $\pmb{\tau}_i$ is now the (equilibrium) steady state of the system when coupled only to the $i$-th bath\footnote{If the dynamics under $\mathcal{L}_{\pmb{H}}^{(i)}$ has more than one fixed point, the $\pmb{\tau}_i$ in this expression could be any them. Here, we shall assume that there is only one stationary state.}. In particular, the above still holds if we replace $\pmb{\varrho}_S(t)$ by the non-equilibrium steady state $\pmb{\tau}$; namely,
\begin{equation*}
\tr{\left[\left(\mathcal{L}_{\pmb{H}}^{(i)}\,\pmb{\tau}\right)\,\log{\pmb{\tau}} - \left(\mathcal{L}_{\pmb{H}}^{(i)}\,\pmb{\tau}\right)\,\log{\pmb{\tau}_i}\right]} \leq 0,
\end{equation*}
where we have used the fact that the dynamics is trace-preserving, i.e., $\tr\,d\pmb{\tau}/dt = 0$. We can now sum over all baths to obtain
\begin{equation}\label{eq:heat-current-intermediate}
-\sum\nolimits_i\tr{\left[ \left(\mathcal{L}_{\pmb{H}}^{(i)}\,\pmb{\tau}\right)\,\log{\pmb{\tau}_i}\right]} \leq 0.
\end{equation}
To get here, we have invoked the stationarity of $\pmb{\tau}$---that is, $ -i[\pmb{H}+\Delta\pmb{H},\pmb{\tau}] + \sum\nolimits_i \mathcal{L}_{\pmb{H}}^{(i)}\,\pmb{\tau} = 0 $---and the linearity of the trace. 

Also, the GKLS generators have the local equilibrium state (at the corresponding temperature) as fixed point. Hence, $ \pmb{\tau}_i = \pazocal{Z}_i^{-1} e^{-\pmb{H}/T_i} $ ($\pazocal{Z}_i = \tr{e^{-\pmb{H}/T_i}}$), so that $\log{\pmb{\tau}_i} = -\pmb{H}/T_i - \log{\pazocal{Z}_i}$. Plugging this back into Eq.~\eqref{eq:heat-current-intermediate} and exploiting again the preservation of the trace finally gives
\begin{equation*}
    \sum\nolimits_i \frac{\tr{\left[\pmb{H}\,\mathcal{L}_{\pmb{H}}^{(i)}\,\pmb{\tau}\right]}}{T_i} \leq 0,
\end{equation*}
which coincides with Eq.~\eqref{eq:clausius-inequality} upon identifying
\begin{equation}\label{eq:heat-current}
    \dot{\mathcal{Q}}_i = \tr{\left[\pmb{H}\,\mathcal{L}_{\pmb{H}}^{(i)}\,\pmb{\tau}\right]}.
\end{equation}
This is a very natural definition for the $ i $-th stationary heat current: namely, it is the asymptotic rate of change of $\langle\pmb{H}\rangle$, due to the dissipative coupling to bath $ i $.

\section{The reorganised master equation}\label{sec:reorganised-master-equation}

\blue{From now on, we shall focus on bosonic baths with Hamiltonian 
\begin{equation}\label{eq:h_bath}
    \pmb H_{B_i} = \sum\nolimits_\mu \frac12\omega_{\mu\,i}^2\,m_{\mu\,i}\,\pmb x_{\mu\,i}^2  + \frac{1}{2\,m_{\mu\,i}}\,\pmb p_{\mu\,i}^2,
\end{equation}
and assume that the $\pmb{B}_i$ are linear operators, e.g., $ \zeta_i\,\pmb{B}_i = \sum\nolimits_\mu g_{\mu\,i}\,\pmb{x}_{\mu\,i} $, in the quadratures of the environmental modes. Such linear baths are widely used to model photonic or phononic environments and, more generally, often serve as a phenomenological template for unspecified baths.}

So far, we have not yet specified which reference Hamiltonian $ \pmb{H} $ to use. It would appear most natural to choose the microscopic Hamiltonian $\pmb{H}_S$ of the open system. \blue{Here, by $\pmb{H}_S$ we mean the bare Hamiltonian $ \pmb{H}_S^{(0)} $ of the system in isolation, plus any corrections that may arise from the microscopic system--bath(s) coupling mechanism \cite{caldeira1983tunnelling,correa2023}; these corrections are often referred-to as counter terms.} However, setting $\pmb{H} = \pmb{H}_S$ is not necessarily a good choice \cite{winczewski2021,lobejko202,correa2023}. Indeed, the classical (or high-temperature) limit $ \pmb{\tau}_\text{cl} $ for the steady state of an arbitrary open system coupled to \textit{one} linear heat bath at temperature $T$ is given by \cite{cerisola2022,cresser2021,timofeev2022,correa2023}
\begin{equation}\label{eq:general-high-T-mean-force}
	\pmb{\tau}_\text{cl} \simeq \frac{e^{-(\pmb{H}_S - Q\,\pmb{S}^2)/T}}{\tr{e^{-(\pmb{H}_S - Q\,\pmb{S}^2)/T}}}.
\end{equation}
$ Q $ is the \textit{reorganisation energy}, which is of second--order in the coupling strength $\zeta$, and captures the distortion of the system potential due to the weak---but finite---coupling to the bath. 
In contrast, a master equation in GKLS form with $ \pmb{H} = \pmb{H}_S $, i.e.,
\begin{equation}\label{eq:conventional}
    \frac{d\pmb{\varrho}_S}{dt} = -i[\pmb{H}_S-\Delta\pmb{H}_S,\pmb{\varrho}_S] + \mathcal{L}_{\pmb{H}_S}\,\pmb{\varrho}_S,
\end{equation}
would drive the system to the wrong steady state \cite{spohn1978}; namely, 
\begin{equation*}
	\pmb{\pi} = \frac{e^{-\pmb{H}_S/T}}{\tr{e^{-\pmb{H}_S/T}}}.
\end{equation*}

In turn, a GKLS master equation like
\begin{equation}\label{eq:reorganised-master-equation}
    \frac{d\pmb{\varrho}_S}{dt} = -i[\pmb{H}_S-Q\,\pmb{S}^2,\pmb{\varrho}_S] + \mathcal{L}_{\pmb{H}_S-Q\,\pmb{S}^2}\,\pmb{\varrho}_S,
\end{equation}
does asymptotically force the system into the classical limit \eqref{eq:general-high-T-mean-force} \cite{winczewski2021}. While it may seem that \eqref{eq:reorganised-master-equation} is a crude manipulation of \eqref{eq:conventional} with dubious validity, it does follow from a rigorous microscopic derivation to $\pazocal{O}({\zeta^2})$ \cite{correa2023}. The trick to obtain it is to cast $ \pmb{H}_S = (\pmb{H}_S - Q\,\pmb{S}^2) + Q\,\pmb{S}^2 \coloneqq \pmb{H}_S^{(R)} + Q\,\pmb{S}^2 $ and treat the second term perturbatively, recalling that it is already of second order in $\zeta$. If, in addition, one assumes that the bath has a broad power spectrum, extending over to sufficiently high frequencies (i.e., that one works in the adiabatic regime), then Eq.~\eqref{eq:reorganised-master-equation} follows naturally \cite{correa2023} \footnote{It is important to note that the non-secular contributions to the Lamb shift term appearing throughout the microscopic derivation of Eq.~\eqref{eq:reorganised-master-equation} are instrumental in arriving at its final form. See Ref.~\cite{correa2023} for details.}. We shall refer to it as the \textit{reorganised master equation}. Its derivation is outlined in Appendix~\ref{app:derivation}. This equation does not only succeed at predicting the correct high-temperature steady state, but it accurately tracks the transient oscillations, where it is valid. 

In principle, one could subtract and add \textit{any} Hamiltonian correction to $\pmb{H}_S = (\pmb{H}_S - \delta\pmb{H}) + \delta\pmb{H}$ and proceed in the same way, provided that $\delta\pmb{H}$ is $\pazocal{O}(\zeta^2)$ \cite{thingna2012,lobejko202,timofeev2022}. One should expect the resulting master equation, 
\begin{equation}\label{eq:arbitrarily-reorganised-me}
    \frac{d\pmb{\varrho}_S}{dt} = -i[\pmb{H}_S-\Delta(\pmb{H}_S-\delta\pmb{H}),\pmb{\varrho}_S] + \mathcal{L}_{\pmb{H}_S-\delta\pmb{H}}\,\pmb{\varrho}_S,
\end{equation}
to roughly agree with Eqs.~\eqref{eq:reorganised-master-equation} or \eqref{eq:conventional} for sufficiently small $\zeta$. That is, Eq.~\eqref{eq:reorganised-master-equation} is one among many possible second-order GKLS master equations---its chief advantage is that it asymptotically drives the system to the correct classical limit. Note as well that the Lamb shift term $\Delta(\pmb{H}_S-\delta\pmb{H})$ for the modified Hamiltonian needs not commute with $ \pmb{H}_S $. As a result, and in contrast to Eq.~\eqref{eq:reorganised-master-equation}, the steady state of \eqref{eq:arbitrarily-reorganised-me} will generally contain coherences in the reference basis for dissipation \cite{winczewski2021}.

\subsection{Conflicting definitions of steady-state heat currents}\label{sec:conflicting-definitions-heat-currents}

\begin{figure}[t]
\centering
\includegraphics[width=0.48\textwidth]{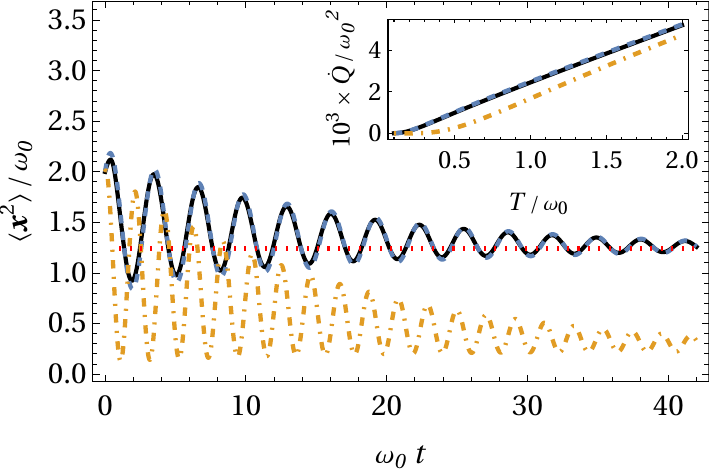}
\caption{\textbf{Dynamics of an oscillator in contact with two heat baths.} The position variance $\textstyle{\langle \pmb{x}^2(t) \rangle}$ as predicted by the reorganised master equation \eqref{eq:reorganised-master-equation} (dashed blue) tracks the exact solution (solid black). In contrast, choosing the full physical Hamiltonian $\pmb{H}_S$ as reference (cf. Eq.~\eqref{eq:conventional}) performs much worse (dot-dashed orange): neither the steady state value (dotted red) nor the transient oscillations are well captured. The inset, shows a similar behaviour for the stationary heat currents (the color coding is the same). Note how the reorganised heat current $\dot{\mathcal{Q}}\big(\pmb{H}_S^{(R)}\big)$ from Eq.~\eqref{eq:reorganised-heat-current} works well even at low temperatures, while the $\dot{\mathcal{Q}}(\pmb{H}_S)$ in Eq.~\eqref{eq:conventional-heat-current} only works at high temperatures. A large reorganisation energy $ Q $ was chosen to better appreciate the differences. The parameters are $\blue{\lambda_1 = 2\,\lambda_2} = 0.01$, $\Lambda_1/\omega_0 = \Lambda_2/\omega_0 = 100$, $T_1/\omega_0 = 1$, and $T_2/\omega_0 = 1.5$ ($\omega_0 = 1$). \blue{This results in $ \sum_i Q_i/\omega_0^2 = 3/4$.} In the inset $ T = T_1 = T_2/1.2$.}
\label{fig1}
\centering
\end{figure}

\blue{As already advanced, setting $ \pmb{H} = \pmb{H}_S - Q\,\pmb{S}^2 $ leads to the reorganised master equation \eqref{eq:reorganised-master-equation} in the adiabatic limit (cf. Appendix~\ref{app:derivation}). In cases in which the bare Hamiltonian $\pmb{H}_S^{(0)}$ must be supplemented with a counter term when coupled to the environment, i.e., $ \pmb{H}_S = \pmb{H}_S^{(0)} + Q\,\pmb{S}^2 $, the aforementioned choice of reference would result in $\pmb{H} = \pmb{H}_S - Q\,\pmb{S}^2 = \pmb{H}_S^{(0)} $ \cite{winczewski2021,correa2023}. Notably, such counter terms appear in the common problem of a system coupled to the electromagnetic field \cite{caldeira1983tunnelling}.} If one deals with multiple baths, the corresponding stationary heat currents would then read
\begin{equation}\label{eq:reorganised-heat-current}
    \dot{\mathcal{Q}}_i\big(\pmb{H}_S^{(R)}\big) = \tr{\big[ \pmb{H}_S^{(R)}\mathcal{L}_{\pmb{H}_S^{(R)}}\,\pmb{\tau}_{\pmb{H}_S^{(R)}}\big]}
\end{equation}
in terms of the reorganised Hamiltonian $ \pmb{H}_S^{(R)} = \pmb{H}_S - \sum\nolimits_i Q_i\,\pmb{S}_i^2 $. Note that we have made the dependence of all objects with the choice of relaxation basis $ \pmb{H} = \pmb{H}_S^{(R)} $ explicit, for clarity.

In contrast, using $ \pmb{H} = \pmb{H}_S $ as the reference, results in Eq.~\eqref{eq:conventional} instead, and the corresponding steady-state heat currents are of the form 
\begin{equation}\label{eq:conventional-heat-current}
    \dot{\mathcal{Q}}_i(\pmb{H}_S) = \tr{\left[ \pmb{H}_S\,\mathcal{L}_{\pmb{H}_S}\,\pmb{\tau}_{\pmb{H}_S} \right]}.
\end{equation}

At first, one might expect Eqs.~\eqref{eq:reorganised-heat-current} and \eqref{eq:conventional-heat-current} to yield similar results, especially since
\begin{equation*}
    \dot{\mathcal{Q}}_i = \tr{\big[ \pmb{H}_S^{(R)}\,\mathcal{L}_{\pmb{H}_S^{(R)}}\,\pmb{\tau}_{\pmb{H}_S^{(R)}}\big]} = \tr{\big[ \pmb{H}_S\,\mathcal{L}_{\pmb{H}_S^{(R)}}\,\pmb{\tau}_{\pmb{H}_S^{(R)}}\big]} + \pazocal{O}(\zeta_i^4).
\end{equation*}
However, as we shall illustrate next, Eq.~\eqref{eq:conventional-heat-current} may deviate substantially from the exact heat current, especially at low temperatures, while the `reorganised heat current' \eqref{eq:reorganised-heat-current} closely follows the exact results. This highlights the importance of a careful choice of master equation when modelling nanoscale heat devices and, importantly, it also shows that the reorganised master equation framework, which correctly describes equilibration in the classical limit \cite{winczewski2021,correa2023}, continues to be accurate in non-equilibrium situations.

\section{Benchmarking}\label{sec:examples}

\begin{figure*}[t]
\centering
\includegraphics[width=0.96\textwidth]{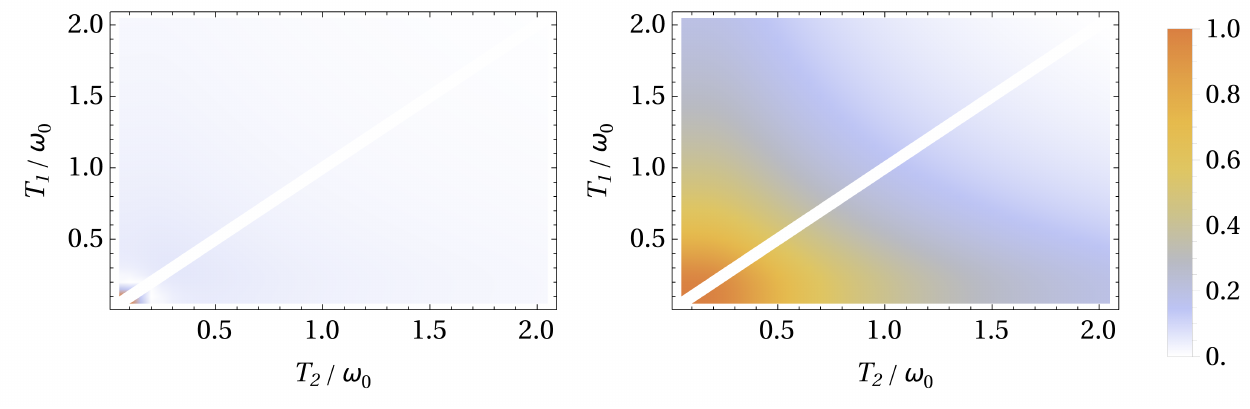}
\caption{\textbf{Benchmarking the heat currents of the oscillator.} Absolute relative difference between the heat current predicted by a master equation and the exact expression \eqref{eq:hc-ex-osc}, for varying temperatures of the baths. The current $\dot{\mathcal{Q}}_i\big(\pmb{H}_S^{(R)}\big)$ from the reorganised master equation \eqref{eq:reorganised-master-equation} (left) proves accurate, except at very low temperatures. In contrast, $\dot{\mathcal{Q}}(\pmb{H}_S)$ (right) starts to break down already at intermediate temperatures. The parameters are the same as in Fig.~\ref{fig1}. The diagonals have been left out to avoid division by zero, as no energy flows if $ T_1 = T_2 $.}
\label{fig2}
\centering
\end{figure*}

In what follows, we compare the accuracy of the predictions of Eqs.~\eqref{eq:reorganised-master-equation} and \eqref{eq:conventional} by benchmarking them against the exact solutions for two paradigmatic open-system models---the damped quantum harmonic oscillator and the spin--boson model. We pay particular attention to the resulting steady-state heat currents; namely, $ \dot{\mathcal{Q}}\big(\pmb{H}_S^{(R)}\big) $ and $ \dot{\mathcal{Q}}(\pmb{H}_S) $.

We consider only two baths, so that $i \in \{1,2\}$. Going back to Eq.~\eqref{eq:H_int} and the couplings $ g_{\mu\,i} $ between the system and each of the environmental modes, these are encapsulated in the spectral densities $J_i(\omega) = \pi\,\sum\nolimits_\mu g_{\mu\,i}^2/(2 m_{\mu\,i} \omega_{\mu\,i})\,\delta(\omega - \omega_{\mu\,i})$ \cite{weiss1999, bp}. We choose the common Ohmic--algebraic profile, given by 
\begin{equation}\label{eq:sd}
	J_i(\omega) = \frac{\lambda_i\,\omega}{1 + (\omega/\Lambda_i)^2}.    
\end{equation}
Here $ \Lambda_i $ is the high-frequency cutoff and the parameter $ \lambda_i $ is the dissipation strength. The reorganisation energies may be computed from the corresponding spectral densities as 
\begin{equation}
\blue{Q_i = \frac{1}{\pi}\int_{0}^\infty d\omega\,\frac{J_i(\omega)}{\omega}.}
\end{equation}
For instance, $ Q_i = \lambda_i\,\Lambda_i/2$ for our particular choice \eqref{eq:sd}. In turn, the dissipation rates are
\begin{align}\label{eq:decay-rates}
    \gamma_i(\omega) = 2\,J_i(\omega)\,(n_{T_i}+1),
\end{align}
where $n_{T_i} = (e^{\omega/T_i}-1)^{-1}$.

We still need to define the Lamb-shift term $ \Delta\pmb{H}_S $ in Eq.~\eqref{eq:conventional}. For a GKLS dissipator, this is given by
\begin{equation}\label{eq:lamb-shift-term}
    \Delta\pmb{H}_S = \sum\nolimits_\omega S_i(\omega)\,{\pmb{A}^{(i)}_\omega}^\dagger\pmb{A}_\omega^{(i)},
\end{equation}
where the coefficients are $S_i(\omega) = -\pazocal{H}\,[J(\omega)\,( n_{T_i}(\omega)+1 )]$. The Hilbert transform is $ \pazocal{H}[f(\omega)] \coloneqq \frac{1}{\pi}\dashint_{-\infty}^\infty d\omega' f(\omega')/(\omega'-\omega) $ and $\dashint$ denotes principal value integral.

Finally, we must define the exact steady-state heat currents. We assume a factorised initial condition for the global system--bath state, of the form $ \pmb{\varrho}(0) = \pmb{\varrho}_S(0)\bigotimes_i\pmb{\pi}_i$. That is, the system may be in an arbitrary state, while the baths are in thermal equilibrium at their temperatures $ T_i $. However, we make no assumptions about the correlations built up in the subsequent dynamics. Note that $\langle \pmb{B}_i(t) \rangle = 0$, where $\pmb{B}_i(t)$ is the coupling operator from the $i$-th bath in the Heisenberg picture and the angle brackets denote averaging with respect to $\pmb{\varrho}(0)$. Also $[\pmb{\pi}_i,\pmb{H}_{B_i}] = 0$. Hence, energy balance in the steady state requires that
\begin{equation*}
    \frac{d}{dt}\langle \pmb{H}_S \rangle = i \sum\nolimits_j\left\langle \left[\zeta_j\,\pmb S_j(t) \otimes \pmb B_j(t),\pmb H_S(t) \otimes \pmb{\mathbbm{1}}\right]\right\rangle = 0,
\end{equation*}
so that the exact heat currents may be identified as
\begin{equation}\label{eq:hc-ex}
	\dot{\mathcal{Q}}_j^\text{\,ex} = i \left\langle \left[\zeta_j\,\pmb S_j(t) \otimes \pmb B_j(t),\pmb H_S(t) \otimes \pmb{\mathbbm{1}}\right]\right\rangle.
\end{equation}
\blue{Equivalently, the steady state heat flow into the system can be calculated as minus the change of energy of the baths \cite{alipour2016correlations}.}
To avoid subtleties in defining the heat currents, we study interactions fulfilling $\left[\pmb{S}_1,\pmb{S}_2\right] = 0$. This is trivially fulfilled when the various sites of a multipartite system couple locally to their own independent heat bath, e.g., a spin chain coupled to one bath at each end.

\subsection{Damped quantum harmonic oscillator}

Our first example is the damped quantum harmonic oscillator \cite{caldeira1983path, caldeira1983tunnelling, hanggi2005, lampo2019, cresser2023} with bare Hamiltonian 
\begin{equation*}
	\pmb H_S^{(0)} = \frac{1}{2} \omega_0^2\,\pmb x^2 + \frac{1}{2}\,\pmb p^2.
\end{equation*}
Here we set $m=1$. The coupling operator will be, in this case, $\pmb S_{1,2} = \pmb x$. We assume that the microscopic Hamiltonian does include a counter term. For this choice of coupling operator, the counter term would shift the frequency to $\omega_R^2 = \omega_0^2 + \sum\nolimits_i\lambda_i\,\Lambda_i$. As the total Hamiltonian is quadratic, this model is exactly solvable and thus, a natural candidate to benchmark approximate methods. 

In particular, the steady-state heat currents are \cite{freitas2014}
\begin{multline}\label{eq:hc-ex-osc}
	\mathcal{\dot Q}^\text{\,ex}_i = \frac{4}{\pi} \int_0^\infty d\omega \, \omega \, J_1(\omega)\, J_2(\omega)\,\abs{\hat{g}(i \omega)}^2 \\\times \left( \coth{\frac{\omega}{2 T_i}}-\coth{\frac{\omega}{2 T_j}}\right),
\end{multline}
where $T_j$ is the temperature of the other bath and 
\begin{equation*}
	\hat{g}(s)=\left( s^2 + \omega_R^2  - \widehat{\chi}(s)\right)^{-1}.
\end{equation*}
Here, $\widehat{\chi}(s)$ is the Laplace transform of the \textit{dissipation kernel}
\begin{equation*}
	\chi(t) = \frac{2}{\pi} \int_0^\infty d\omega\, \left[ J_1(\omega) + J_2(\omega) \right] \sin{\omega t}.
\end{equation*}
For our spectral densities, this evaluates to
\begin{equation*}\label{eq:covariance-matrix-general}
	\widehat{\chi}(s) =  \sum\nolimits_i\frac{\lambda_i\, \Lambda_i^2}{s + \Lambda_i}.
\end{equation*} 

When it comes to the master equations, the operators $\pmb{A}_\omega^{(i)} = \frac12(\pmb{x}+\frac{i}{\omega}\pmb{p})$. In the case of the reorganised master equation, the frequencies are $\omega = \pm \omega_0$, while the Bohr frequencies of $ \pmb{H}_S $ are $\omega = \pm \sqrt{\omega_0 + \sum_i \lambda_i\,\Lambda_i}$. These are all the ingredients needed to put formulas \eqref{eq:reorganised-heat-current} and \eqref{eq:conventional-heat-current} to the test.

In Fig.~\ref{fig1} we illustrate how both the dynamics and the stationary heat current current are more accurately captured by the reorganised master equation, particularly at low temperatures. Remarkably, the reorganised master equation accurately tracks the dynamics and current even for moderate reorganisation energies \cite{correa2023}. 

As highlighted in Ref.~\cite{correa2023}, the reorganised framework is accurate in describing the system dynamics, regardless of whether the dissipators are in GKLS form or whether they keep the so-called non-secular terms. The latter case is referred-to as (Born--Markov) Redfield equation \cite{redfield1957}. Importantly, however, Redfield equations do not enjoy the complete positivity of GKLS generators and a relationship of the form \eqref{eq:clausius-inequality} is not guaranteed. They can, however, be used to define heat currents in the form \eqref{eq:heat-current}, provided that the corresponding Lamb-shift correction is absorbed into the dissipators. If the system--bath coupling remains weak, such Redfield currents seem to behave in a thermodynamically consistent manner \cite{strasberg2016}. Indeed, we also computed the Redfield heat currents for both of our examples and verified that they coincide with the GKLS reorganised currents.

In order to study the currents in more detail, in Fig.~\ref{fig2} we plot the absolute relative difference between the heat currents from Eqs.~\eqref{eq:reorganised-heat-current} and \eqref{eq:conventional-heat-current}, and the exact expression \eqref{eq:hc-ex-osc}, i.e.,
\begin{equation*}
	\frac{\abs{\mathcal{\dot Q}(\pmb{H})-\mathcal{\dot Q}^\text{ ex}_i}}{\mathcal{\dot Q}^\text{ ex}_i},
\end{equation*}
for varying temperatures of the baths. We find that setting $\pmb{H} = \pmb{H}_S^{(R)}$ yields very accurate results, except at very low temperatures, while the choice $\pmb{H} = \pmb{H}_S$ leads to much poorer performance, other than at very high temperatures.

\subsection{Spin--boson model}

\begin{figure*}[t]
\centering
\includegraphics[width=0.48\textwidth]{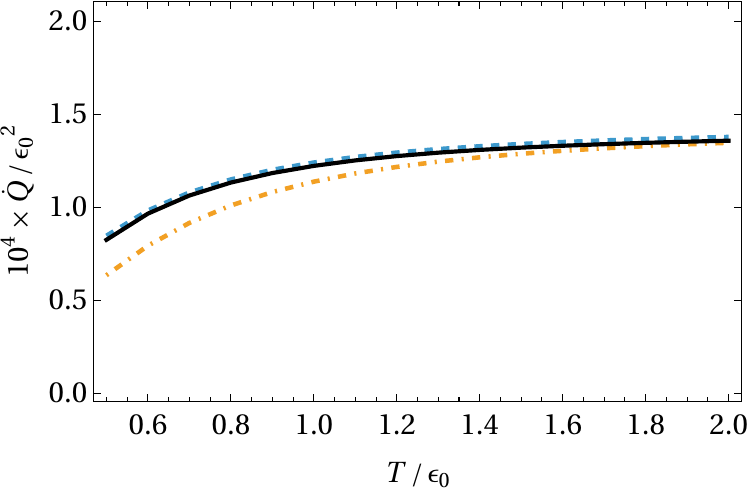}
\includegraphics[width=0.35\textwidth]{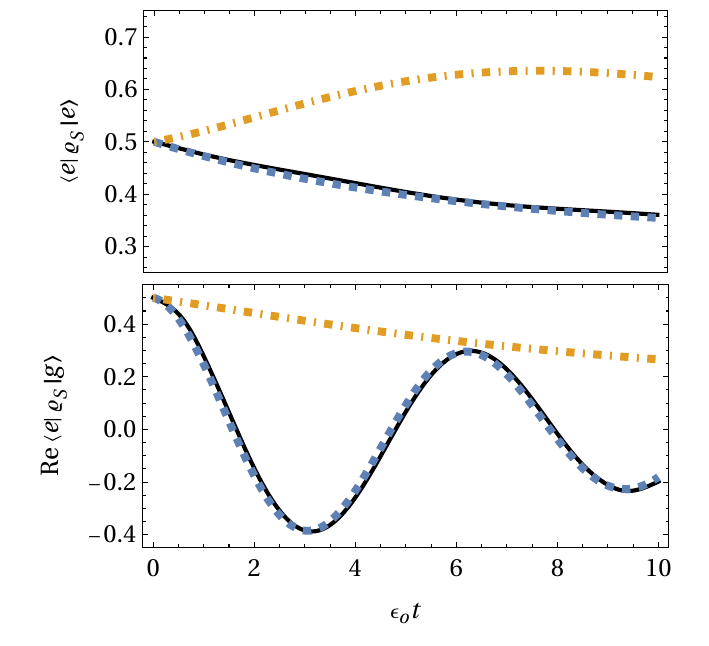}
\caption{\textbf{Heat currents of a spin in contact with two heat baths.} The steady state heat current (left) of a spin in contact with two heat baths at different temperatures as calculated from the reorganised master \eqref{eq:reorganised-heat-current} equation (dashed blue) accurately reproduces the exact value \eqref{eq:hc-ex} (solid black). In contrast, $\dot{\mathcal{Q}}_i(\pmb{H}_S)$ (cf. Eq.~\eqref{eq:conventional-heat-current}) (dot-dashed orange) performs worse, especially at low temperatures. The parameters are $\lambda_1= 3\times 10^{-3}$, $ \lambda_2=1 \times 10^{-3}$, $ \Lambda_1/\epsilon_0 = \Lambda_2/\epsilon_0 = 100$ and $T = T_1 = T_2 / 1.2$ ($\epsilon_0 = 1$). \blue{We have checked that the heat currents $\dot{\mathcal{Q}}_i\big(\pmb{H}_S^{(R)}\big)$ start to deviate from the exact solution as the $ Q_i $ grow larger. Nonetheless, Eq.~\eqref{eq:reorganised-master-equation} continues to yield accurate results even at \textit{moderate} reorganisation energies, as we illustrate in the figure.} In particular, we compare predictions for the excited state population (upper right panel) and real part of the coherence (lower right panel). A maximally coherent initial state was chosen, $\pmb{\varrho}_S(0) = \ket{+}\bra{+}$, with $\ket{+} = \frac{1}{\sqrt{2}}(\ket{g} + \ket{e})$ and $\ket{e}$ ($\ket{g}$) representing the excited (ground) state of the spin-1/2. \blue{Specifically, in the right-hand side panels, $\lambda_1 = 2\,\lambda_2 = 2 \times 10^{-2} $, $\Lambda_1/\epsilon_0 = \Lambda_2/\epsilon_0 = 50$, $T_1/\omega_0=1$, and $T_2/\omega_0=2$, which gives again $\sum_i Q_i/\epsilon_0^2 = 3/4$. The corresponding exact steady state current is $4.49 \times 10^{-3}\,\epsilon_0^2$, while the reorganised master equation yields $4.56 \times 10^{-3}\,\epsilon_0^2$, being off by only $1.6\%$.}} 
\label{fig3}
\centering
\end{figure*}

We now turn to a nonlinear model; namely, the well-known spin--boson model \cite{lambert2019,boudjada2014,yang2014,purkayastha2020,thoss2001,anders2007}, with bare Hamiltonian
\begin{equation}\label{eq:sys-qb}
	\pmb H_S^{(0)} = \frac{1}{2} \epsilon_0\,\pmb \sigma_z,
\end{equation}
and consider the coupling operator
$\pmb S_{1,2} = \pmb{\mathbbm{1}} + \pmb \sigma_x$, where $\pmb\sigma_\alpha$ are Pauli matrices. Just as in the oscillator case, we assume that $ \pmb{H}_S $ contains a counter term, so that, once again, $ \pmb{H}_S^{(R)} = \pmb{H}_S^{(0)} $. Notice that we have ensured that our couplings have non-zero trace; otherwise, the $ Q_i\,\pmb{S}_i^2 $ correction would be proportional to the identity and thus, irrelevant in practice \cite{cresser2021,timofeev2022,cerisola2022}.

To solve for the dynamics and heat currents, we use the numerically exact hierarchical equations of motion approach (HEOM) \cite{tanimura1989, tanimura2020}. In particular, we resort to the QuTiP-BoFiN implementation \cite{lambert2023},  which is integrated in the QuTiP platform \cite{johansson2012, johansson2013}. The numerics require the truncation of two expansions. The first one is the Matsubara expansion of the bath correlation function in an exponential series, where $N_k$ terms are retained. The second one, is the level of the hierarchy for the so-called auxiliary density operators, $N_C$. We have checked our results for convergence in both $N_k$ and $N_C$. In our calculations, we have used the Tanimura terminator \cite{lambert2023} to speed up convergence. The average heat currents in the non-equilibrium steady state can be calculated according to Eq.~\eqref{eq:hc-ex} by using the auxiliary density matrices \cite{tanimura2020}.   

Similarly to Fig.~\ref{fig1}, in Fig.~\ref{fig3} (left panel) we study the steady state heat currents of a spin coupled to two bosonic baths at different temperatures. We find that the predictions of Eq.~\eqref{eq:conventional} break down at intermediate-to-low temperatures, while the reorganised master equation remains accurate in the entire range probed. \blue{Just as in the case of the dissipative harmonic oscillator, in the spin--boson model the quantitative agreement between the exact and the reorganised approach is good at low and even up to moderate reorganisation energies $ Q_i $, as shown in the right-hand side panels of Fig.~\ref{fig3}. In contrast, Eq.~\eqref{eq:conventional} may fail dramatically.}

\section{Conclusions}\label{sec:conclusions}

We have studied the fitness of the reorganised master equation for estimating the stationary heat currents flowing through non-equilibrium open systems. In deriving this equation, the freedom to set the reference basis for dissipative processes is exploited, so as to ensure correct equilibration in the classical limit. \blue{That is, the reorganised master equation attains the high-temperature limit of the exact steady state of the system.} This is achieved by working in the basis of the microscopic system Hamiltonian renormalised by the system--baths couplings. As we have shown, considering two paradigmatic examples, this results in much more accurate estimates of the heat currents, especially at low temperatures. \blue{Interestingly, the quantitative agreement of the heat currents calculated through the reorganised master equation extends even up to moderate reorganisation energies $ Q/\epsilon \sim 1 $ in both models although, as it should be expected the reorganised master equation eventually breaks down at large $ Q/\epsilon $ ($\epsilon$ here is the relevant energy scale of the system).}

Importantly, non-secular (Redfield) reorganised master equations can be brought into the canonical GKLS form via the secular approximation without significant loss in precision, neither in the dynamical description nor in the resulting steady-state heat currents. The same is not true, however, for their `non-reorganised' counterparts---these predict substantially wrong dynamics, as illustrated in both of our examples. The GKLS form is particularly important for quantum thermodynamics, as completely positive generators guarantee thermodynamic consistency.

The manipulations leading to the reorganised master equation may be exploited to derive a whole family of equations. These would all be accurate to the same order in perturbation theory, but can be engineered to have different desirable properties. For instance, one could think of matching the resulting steady state with the true mean-force Gibbs state, rather than with its classical counterpart. This certainly deserves further investigation. Another open question is to clarify the precise relationship between coupling strength $\zeta$, dissipation parameter $ \lambda $, environmental cutoff frequency $ \Lambda $, and reorganisation energy $ Q $, as all of these quantities seem to play a key role in the derivation of perturbative Born--Markov master equations. \blue{Finally, elucidating whether these results extend to fermionic baths is a problem that certainly deserves independent study}.

\acknowledgements

We acknowledge the University of La Laguna (ULL) and the Spanish Ministry of Universities for supporting the \textit{``DQDD Quantum Thermometry Program''}, during which this project started. JG, LAC and DA acknowledge funding by the Ministerio de Ciencia e Innovación and European Union (FEDER) (PID2022-138269NB-I00). JG is supported by a scholarship from CEMPS at the University of Exeter. LAC is supported by a Ram\'{o}n y Cajal fellowship (RYC2021-325804-I), funded by MCIN/AEI/10.13039/501100011033 and “NextGenerationEU”/PRTR. 

\appendix

\onecolumngrid

\color{black}

\section{Derivation of the reorganised master equation}\label{app:derivation}

In order to make the present manuscript more self-contained, we now outline the derivation of the reorganised master equation as put forward in Ref.~\cite{correa2023}. For simplicity, we consider only one bath, although the generalisation to any number of baths is immediate. Let us begin by considering the Redfield equation, which may be derived from the exact Liouville--von Neumann equation under the following assumptions:
\begin{itemize}
    \item weak system--bath coupling $ \zeta $,
    \item comparatively short bath memory defined by the decay of the two-time correlation functions $\langle\tilde{\pmb{B}}_i(t)\tilde{\pmb{B}}_i(t-s)\rangle$,
    \item initially uncorrelated system--bath state, with the bath in thermal equilibrium, i.e., $\pmb{\rho}(0) = \pmb{\varrho}_S(0)\otimes\pmb{\pi}_B$, 
    \item and the further technical requirement that $ \langle \tilde{\pmb{B}}_i(0) \rangle = 0 $ (cf. foonote on page \pageref{foot:first-order}).
\end{itemize}
The resulting second-order master equation takes the form
\begin{equation}\label{eq:true-2ndorder-redfield}
	\frac{d\widetilde{\pmb\varrho}_S}{dt} = - \zeta^2\int_0^\infty ds\,\tr_B\,[\widetilde{\pmb{V}}(t),\,[\widetilde{\pmb{V}}(t-s),\,\widetilde{\pmb\varrho}_S(t)\otimes\pmb{\pi}_B]] + \pazocal{O}(\zeta^3),
\end{equation}
where the tilde denotes interaction picture operators with respect to the uncoupled system-plus-bath Hamiltonian $\pmb{H}_S + \pmb{H}_B $ and $\pmb{H}_\text{int} = \zeta\,\pmb{V} = \pmb{S}\otimes\pmb{B}$. Averages are taken with respect to the initial equilibrium state of the bath. This may be rewritten as
\begin{equation}\label{eq:Redfield-explicit}
	\frac{d\pmb{\varrho}_S}{dt} = -i\,[\pmb{H}_S+\Delta\pmb{H}_\text{full},\pmb{\varrho}_S] + \sum\nolimits_{\,\omega,\omega'}\Gamma_{\omega}\left( \pmb{A}_\omega\,\pmb{\varrho}_S\,\pmb{A}_{\omega'}^\dagger - \frac12\,\{\pmb{A}_{\omega'}^\dagger\,\pmb{A}_\omega,\,\pmb{\varrho}_S\} \right) + \text{h.c.},
\end{equation}
where the complex decay rates $\Gamma_\omega $ are
\begin{equation}\label{eq:complex-rates}
	\Gamma_\omega=\int_0^\infty ds\,e^{i\omega s}\,\langle \widetilde{\pmb{B}}(t)\widetilde{\pmb{B}}(t-s) \rangle_B = \frac12\,\gamma(\omega) + i\,S(\omega).
\end{equation}    
The non-Hermitian operators $ \pmb{A}_\omega $ were defined in Sec.~\ref{sec:heat-currents} and $ \gamma(\omega) $ and $ S(\omega) $ were introduced in Sec.~\ref{sec:examples}. In turn, the full Lamb shift Hamiltonian $ \Delta\pmb{H}_\text{full} $ is given by
\begin{equation}\label{eq:coherent-hamiltonian-explicit}
    \Delta \pmb{H}_\text{full} = -\frac{i}{2}\sum\nolimits_{\omega,\omega'} \Gamma_{\omega}\, \pmb{A}_{\omega'}^\dagger\,\pmb{A}_\omega + \text{h.c.},
\end{equation}
whereas the secular Lamb-shift term $ \Delta\pmb{H} $ appearing in Eq.~\eqref{eq:generic-master-equation} consists of the sum of only the terms with $ \omega = \omega' $. 

The trick to turn Eq.~\eqref{eq:true-2ndorder-redfield} into the reorganised master equation is work with a different free Hamiltonian; namely, $ \pmb{H}_S-Q\,\pmb{S}^2 $, so that $ \pmb{H}_S = (\pmb{H}_S-Q\,\pmb{S}^2) + Q\,\pmb{S}^2 $. Thus, when taking the interaction picture, we may treat $ Q\,\pmb{S}^2 $ as a perturbation to $ (\pmb{H}_S-Q\,\pmb{S}^2) + H_B $, since it is of order $ \zeta^2 $. This results in
\begin{equation}\label{eq:2nd-order-interaction-picture}
    \widetilde{\pmb{V}}(t) = e^{i\,(\pmb{H}_S + \pmb{H}_B)\,t}\,\Hint\,e^{-i\,(\pmb{H}_S + \pmb{H}_B)\,t} = e^{i\,[(\pmb{H}_S-Q\,\pmb{S}^2) + \pmb{H}_B]\,t}\,\Hint\,e^{-i\,[(\pmb{H}_S-Q\,\pmb{S}^2) + \pmb{H}_B]\,t}  + \pazocal{O}(\zeta^4) = \tilde{\pmb{V}}_0(t) + \pazocal{O}(\zeta^4). 
\end{equation}
Plugging this into Eq.~\eqref{eq:true-2ndorder-redfield} gives us 
\begin{equation}\label{eq:vonNeumann-reorganised-redfield}
	\frac{d\widetilde{\pmb\varrho}_S}{dt} = - \zeta^2\int_0^\infty ds\,\tr_B\,[\widetilde{\pmb{V}}_0(t),\,[\widetilde{\pmb{V}}_0(t-s),\,\widetilde{\pmb\varrho}_S(t)\otimes\pmb{\pi}_B]] + \pazocal{O}(\zeta^3),
\end{equation}
which is as accurate as \eqref{eq:true-2ndorder-redfield}. Hence, Eq.~\eqref{eq:Redfield-explicit} may be replaced by
\begin{equation}\label{eq:Redfield-reorganised-explicit}
    \frac{d\pmb{\varrho}_S}{dt} = -i\,\Big[\pmb{H}_S-\frac{i}{2}\sum_{\nu,\nu'} \Gamma_{\nu}\, {\pmb{A}_{\nu'}^{(0)}}^\dagger\,\pmb{A}^{(0)}_\nu,\pmb{\varrho}_S \Big] + \sum_{\nu,\nu'}\Gamma_{\nu}\left( \pmb{A}^{(0)}_\nu\,\pmb{\varrho}_S\,{\pmb{A}_{\nu'}^{(0)}}^\dagger - \frac12\,\{{\pmb{A}_{\nu'}^{(0)}}^\dagger\,\pmb{A}^{(0)}_\nu,\,\pmb{\varrho}_S\} \right) + \text{h.c.}
\end{equation}
without loss of precision. Here, the $ \pmb{A}_{\nu}^{(0)} $ are eigenoperators of the reorganised system Hamiltonian $ \pmb{H}_S^{(R)} = \pmb{H}_S - Q\,\pmb{S}^2 $ and the double summation runs over the set $ \nu $ of the Bohr frequencies of $ \pmb{H}_S^{(R)} $. 

Finally, if all the Bohr frequencies appearing in Eq.~\eqref{eq:Redfield-reorganised-explicit} are such that $ \nu/\Lambda \ll 1 $, one may approximate all $ S(\nu) \simeq S(0) = - Q $ \cite{correa2023}. It is then easy to see that \eqref{eq:Redfield-reorganised-explicit} takes the final form
\begin{multline}\label{eq:Redfield-reorganised-final}
    \frac{d\pmb{\varrho}_S}{dt} \simeq -i[\pmb{H}_S - Q\,\pmb{S}^2,\pmb{\varrho}_S] -\frac14 \big[\sum\nolimits_{\nu,\nu'}\left(\gamma(\nu)-\gamma(\nu')\right)\,{\pmb{A}_{\nu'}^{(0)}}^\dagger\,\pmb{A}^{(0)}_{\nu} ,\pmb\rho_S\big]  \\ + \frac12\,\sum\nolimits_{\nu,\nu'} \left(\gamma(\nu)+\gamma(\nu')\right)\,\left( \pmb{A}^{(0)}_\nu\,\pmb{\varrho}_S\,{\pmb{A}_{\nu'}^{(0)}}^\dagger - \frac12\,\{{\pmb{A}_{\nu'}^{(0)}}^\dagger\,\pmb{A}^{(0)}_\nu,\,\pmb{\varrho}_S\} \right) + \,\text{h.c.}
\end{multline}

The secular approximation consists in averaging out all terms with $ \nu \neq \nu' $ through coarse-graining \cite{bp}, which finally yields our Eq.~\eqref{eq:reorganised-master-equation} from the main text.

\color{black}

\twocolumngrid

\renewcommand{\thefootnote}{\fnsymbol{footnote}}

\bibliographystyle{apsk}
\bibliography{references}

\end{document}